# TerraService.NET: An Introduction to Web Services

Tom Barclay

Jim Gray

Eric Strand

Steve Ekblad

Jeffrey Richter

# TerraService.NET: An Introduction to Web Services

Tom Barclay. Jim Gray, Steve Ekblad, Eric Strand, Jeffrey Richter
{TBarclay, Gray}@microsoft.com
sekblad@itc.nrcs.usda.gov
estrand@synergetics.com
JeffreyR@Wintellect.com

Microsoft Research, 455 Market Street, Suite 1690, San Francisco, CA 94105
http://research.microsoft.com/barc
USDA - NRCS - Information Technology Center, Ft. Collins, CO 80526
http://itc.nrcs.usda.gov
Synergetics Incorporated, Ft. Collins, CO 80524
http://synergetics.com
http://wintellect.com

*Abstract:*

This article explores the design and construction of a geo-spatial Internet web service application from the host web site perspective and from the perspective of an application using the web service. The TerraService.NET web service was added to the popular TerraServer database and web site with no major structural changes to the database. The article discusses web service design, implementation, and deployment concepts and design guidelines. Web services enable applications that aggregate and interact with information and resources from Internet-scale distributed servers. The article presents the design of two USDA applications that interoperate with database and web service resources in Fort Collins Colorado and the TerraService web service located in Tukwila Washington.

## 1   Introduction: TerraServer and TerraService

The Microsoft® TerraServer (http://terraserver.net) web site has been operational since 1998. It stores aerial, satellite, and topographic images of the earth in an SQL database available via the Internet. It is the world's largest online atlas, combining fifteen terabytes of aerial imagery data and 1.5 terabytes of topographic maps from the United States Geological Survey (USGS). Internet browsers provide intuitive spatial and text interfaces to the data. Users need no special hardware, software, or knowledge to locate and browse imagery.

Although it is a popular web site, many applications have wanted programmatic access to the TerraServer database so that they could incorporate TerraServer data directly without "screen scraping" TerraServer HTML pages. This article describes the *TerraService* a web service that provides programmatic access to the TerraServer database (http://TerraService.Net).

The TerraService was our first production web service. Initially we were confused about what a web service is, how to design a web service, and what tools were needed to build and deploy a web service.

This article answers these introductory questions by example rather than being an abstract discussion of issues – it is a bottom-up tour of the design and construction of real web services and the applications that use them.

## 2   What is a Web Service?

A web service is a web site intended for use by computer programs instead of by human beings. Prior to the invention of web services, most web sites contained documents written in HTML and linked together by anchor tags. A web browser was required to decode the HTML and display it on a computer monitor. An HTML document may reference image files or contain JavaScript functions to give the illusion of an active document or program – but the goal is to create a display image for a person to view and interact with.

An application that integrates data from several web sources wants the semantic content of each web page. Many programs have been written that parse HTML documents to extract the essential information from the interleaved formatting and factual data. This *screen scraping* has had mixed success. XML was created in part to simplify the programming task of generating, exchanging, and parsing semi-structured data.

The World Wide Web Consortium (W3C) has defined message formats and protocols to call and pass parameters to an XML page server. Most applications require query string parameters or specific form fields be filled in with appropriate values in order to generate the HTML or XML document. The W3C developed the Simple Object Access Protocol (SOAP) and related standards that extend XML so that computer programs can easily pass parameters to server applications, and then receive and understand the returned semi-structured XML data document.

More specifically, the SOAP specification is divided into four parts:

1. The SOAP envelope construct defines an overall framework for expressing what is in a message, who should deal with it, whether it is optional or mandatory, and how to signal errors.
2. The SOAP binding framework defines an abstract framework for exchanging SOAP envelopes between

peers using an underlying protocol for transport. The SOAP HTTP binding (SOAP in HTTP) defines a concrete instance of a binding to the HTTP protocol.
3. The SOAP encoding rules defines a serialization mechanism that can be used to exchange instances of application-defined data, arrays, and compound types.
4. The SOAP RPC representation defines a convention that can be used to represent remote procedure calls and responses.

SOAP is augmented with an interface definition language (IDL) called Web Services Definition Language (WSDL). Each web service publishes its interface as an XML document that completely specifies the service's request/response interface so that clients and client tools can automatically bind to the web service. The TerraService WSDL interface document is published at (http://terraservice.net/TerraService.asmx?wsdl).

SOAP has been compared to Microsoft's Distributed Component Object Model (DCOM) and the Object Management Group [OMG] Common Object Request Broker Architecture (CORBA). Like DCOM and CORBA, SOAP provides the framework or "plumbing" to enable two cooperating software programs to inter-operate, but SOAP allows cooperation over the public Internet using vendor-neutral XML-based protocols. Software vendors and Internet consortia are rapidly introducing products and tools that support the SOAP specification and simplify the lives of the programmers on many platforms. Microsoft, IBM, HP, Microsoft, and Apache all have major efforts underway to support SOAP and XML web services.

Web service clients are computer programs that invoke the web service methods. The clients and servers are typically built with integrated program development environments such as Microsoft's Visual Studio .NET that hide all the details of SOAP and XML. The programmer publishes a web service just as they would publish a class implemented in a local DLL file. The tool generates the WSDL and installs all the programs on a web server.

Client programs accesses the web service just the same way they access any other object class. The tool reads the WSDL and translates the class definitions into the programmer's language (C++, Java, C#, Visual Basic, VBscript, Jscript, or any other language.) The tools build the proxy and stub code required to communicate between the client and the remote web service. Thus when a program invokes a published web service method, it is really calling a method implemented by Visual Studio .NET that handles the SOAP/XML encoding/decoding and transmission to the remote web service. The runtime builds a SOAP request document, sends it to the remote web service via HTTP or native SOAP, waits for a response, decodes the returned SOAP/XML document, and formats the returned result in the binary format familiar to the program. For example, the following C# program fragment fetches demographic information about San Francisco from the TerraService:

```
TerraService ts   = new TerraService();
Place place       = new Place();
place.City        = "San Francisco";
place.State       = "California";
place.Country     = "United States of America";
PlaceFacts[] pfs  = ts.GetPlaceFacts(place);
```

This client application will work on any platform (Windows, Macintosh, UNIX, etc.,) running on any computer that has access to the Internet.

In summary, a web service allows a client application to access methods published as a remote public class via Internet protocols over a wide-area, typically public network. Because it is encoded as an XML document, data is easily exchanged between computing systems with incompatible architectures and incompatible data formats. WSDL completely describes the web service interface, and SOAP headers and documents completely describe the data types of parameters, returned data structures, and exceptions. This enables programming environments such as Visual Studio.NET to do all the "heavy lifting" for both the web service consumer and publisher. The publisher simply implements a class with public methods and declares them to be a *Web Method*. A consumer references the published web service class as they would a local or system class and invokes the public methods. The programming environment uses the SOAP system to perform all the RPC and data encoding/decoding operations.

## 3 How to Design a Web Service

Designing a web service is similar to designing any client/server application where the two communicating parties are on separate nodes. The standard object-oriented issues of encapsulation, polymorphism, and inheritance are unchanged in this context. The web service design issues are similar to designing client/server systems using DCOM or CORBA.

This section discusses the generic issues of:
- **Security**: If your service is not public you must authenticate users and implement an access control policy.
- **Atomicity**: Services should leave the web service state in a consistent state.
- **Disconnected Services**: To support disconnected clients SOAP requests can be asynchronous or queued.
- **Granularity**: "1-to-10KB in 1 second" rule of thumb.
- **Stateful vs. Stateless Services**: where possible make the web service invocations stateless so that all client context is passed as a parameter or is stored in the database.

### 3.1 Security

There are many ways to secure and authenticate access to web service methods. By default, web service methods are accessed using anonymous access, the least secure access method. This effectively means that any application or user can execute the web service method. Anonymous access doesn't even require the user to identify themselves.

Today, web services security has not been fully standardized – a collection of standards are being developed by cooperating groups. The .NET framework has a sophisticated security model that offers a spectrum from anonymous access to strong-encryption and authentication based on certificates. Web service security design is a rapidly evolving area worthy of its own article.

TerraService uses anonymous access because all the information accessible by the TerraService is public domain and provided without charge and all the methods are read-only. Therefore, we did not implement an authentication and security protocol.

### 3.2 Atomicity

Web services that make changes to a website's state should be atomic: that is they should either do all of their intended action, or they should do nothing and raise an exception.

Database transactions are the standard technique for implementing this contract. The service begins a transaction and then performs the operations. If all goes well, the service commits the transaction and the database changes are made durable. If there is a problem, the method catches the exception and aborts the transaction causing the state to rollback to the state before the call. The method then returns an error.

Since the TerraService is a stateless and read-only web service, we did not face any of these issues.

### 3.3 Disconnected Operation

The Internet is a loosely connected distributed system. Where possible, applications should use asynchronous web service invocations. If the client is disconnected, then client requests can be queued, and then batched to the service when the client is reconnected. Anyone who uses eMail or a PDA is familiar with these design issues.

All web services have both a synchronous and an asynchronous interface. The asynchronous interface has a *begin* method to invoke the service and an *end* method to poll for the invocation status and fetch the result. There is even a *callBack* mechanism to avoid polling.

The TerraService described illustrates some of these issues. Some clients connect to the service; download a subset of the data; and then use that data in a "disconnected mode". The .NET concept of dataset allows applications to download parts of a database, operate on the data offline, and then propagate updates back to the web service as *updategrams*. Using this queued or asynchronous processing is often essential for the usability of mobile or Internet applications.

### 3.4 Web server granularity: The 1-to-10KB in 1 second rule of thumb

Good client/server implementations pay careful attention to the roundtrip costs over the network between the client and the server. Web service designers need to be even more sensitive to the roundtrip overhead for several reasons. First, the communication will probably be over a wide-area network with relatively low bandwidth and high latency. Second the transmission may be passed through intermediate systems such as firewalls, proxies, routers, and gateways. And third, the data will have to be converted to XML at the source and back to native formats at the destination. So, the cost, latency, and jitter of an SOAP invocation are substantially larger than the costs for a LAN-based DCOM or CORBA invocation.

These overheads are worth the benefits. Publishers can now easily publish data to the entire Internet. Application designers can now design and build wide-area distributed applications that interoperate among a multitude of platforms.

The application designer cannot control the performance the Internet or the distance between Client and Server and so must plan for high latency and slow data rate. She must find a balance between method execution time (method *granularity*), and the amount of data transferred over the Internet. Both parameters have a huge impact on the performance of the application. The designer must find the sweet spot between

- *too simple methods* that require the caller to access the server to frequently (too many roundtrips) and
- *too complicated methods* that take too long to execute on the server or return too much data to be swallowed at once.

We use a "1-to-10KB in 1 second" rule of thumb. We believe that most web service methods should take less than **1 second** to execute and ideally return between **a thousand and ten thousand bytes of data**. Of course there are always valid cases where a method might need more than a second to execute, or where sub-second response is required. But remember, calling a web service method is expensive. You will be putting the client machine through several process transitions to call the web service; the message will traverse firewalls, routers, and web servers before it is dispatched to the server; and, the reply with retrace that path. Processing the request will likely be a small part of the entire invocation. Thus you want to make sure the **value of the result** was worth all the overhead of the call.

The result returned to the client should also be at most a few KB if possible. Again, there are plenty of valid instances where the size of the result will be less than 1KB or greater than 10KB. But optimal web service methods return results in this range. There are three major activities in sending results back to the client. The server must convert from the web service platform's native data type to XML and then wrap the result in a SOAP envelope. This increases the size of the data to be returned by 2x or more [1]. The swollen data must be transmitted to the client, typically in 1.5KB packets (Internet Message Transfer Units (MTUs)). Then the client has to parse the XML and convert it to the client's native data types. Thus, you want to keep the amount of data transmitted and converted to something that can be handled reasonably quickly.

### 3.5 Stateful versus Stateless

HTTP and SOAP are inherently stateless -- SOAP methods expect parameters containing all the context information needed to perform the function. A web method that needs context information from previous invocations must develop its own techniques for storing application or end user context information.

The .NET Framework provides several mechanisms for web service applications to maintain state between method invocations. However, maintaining state can be costly and error-prone depending on web server configuration, proxy servers, firewalls, etc. Therefore, it is generally desirable to implement stateless web service methods whenever possible.

All TerraService web methods are stateless. Context information required by each method is passed to the method through the calling parameters. The results of the method are passed back to the client application and no context information is retained within the web service.

## 4 Web Services in Web Servers

Web servers like Apache and Microsoft's IIS (Internet Information Server) handle all port 80 tcp/ip requests. SOAP requests arrive at port 80 either as HTTP GET or POST requests or as native SOAP messages. The web server converts these SOAP requests into local invocations -- the invoked method is unaware that SOAP is even involved.

Web servers are effectively big switches that de-serialize and dispatch each request to the software that understands how to process that request. The web server then returns the resulting data back to the client.

IIS dispatches in four styles (1) a file system fetch when an HTML, JPEG, GIF or other simple file is requested, (2) a DLL that implements the ISAPI protocol for communication (similar to CGI in Apache), (3) an Active Server Page (ASP) script, and (4) with the .NET Framework the ASP.NET [2] execution environment.

IIS web service invocations are routed to ASP.NET. The file extension ".ASMX" is used to identify the request as a web service in the .NET environment. An ASMX file is a simple text file that tells the ASP.NET run-time where to locate the .NET Class that implements the web service. Typically the .NET Class has the same name as the .ASMX file, but with the .DLL file extension. It is also located in the Bin sub-directory beneath the .ASMX file.

Parameter values travel in the SOAP message tagged by XML parameter tags. Using HTTP GET style invocation, query string parameters follow the web service method URL. To invoke a web service method, IIS dispatches to ASP.NET reads the ASMX file to find the DLL or EXE to load, loads it into memory if it hasn't done so already. It then deserializes the parameters from the SOAP-XML or from the URL and then calls the specified method with these parameters.

The method must exist with the same name (case doesn't matter), and have parameters with the same name and type as those found on the request. ASP.NET returns a SOAP exception message if the caller doesn't get this right.

### 4.1 What is a Web Service to the Programmer?

Toolkits make it easy to build web services. They provide templates and wizards that generate proxies on the client and stubs on the server that hide the invocation details mentioned above and make it trivial to construct a web service class.

In the Visual Studio.NET framework, a *web service* is a *class* that inherits from the System.Web.Services.WebService class. You *annotate* your *public* methods as a *Web Method*. The ASP.NET runtime does the rest: it publishes and invokes these methods. The following C# fragment gives a hint of the programming style.

```
[WebService(Namespace="http://you.org/")]
public class
  yourService : System.Web.Services.WebService {
    [WebMethod]
     public string echo(string name) {
         return name;}
    }
```

The client invokes this method as:
```
yourService ys = new yourService();
string ans  = ts.echo("Hello World!");
```

All parameters are marshaled into an XML SOAP message. The framework provides marshaling for the standard datatypes (numbers, strings, dates, arrays, lists …) but more advanced constructors like dataset have not yet been standardized by the W3C. In essence, you can pass any object and return any object type that has a standardized serializer. Realistically, only atomic-data and SOAP-specified constructors like *array* and *struct* can be marshaled over the wire in a vendor-neutral way. A Class reference is an example of something that cannot be passed there is currently no *open* way to transmit a class over the wire to a different platform. If you are willing to restrict yourself to .Net clients, then you can pass a much richer set of objects. Indeed, in .Net each class can provide its own serialize/deserializes interface - but this will not work outside the .Net universe.

---

[1] Compression schemes like Xmill can reduce this by 10x or more but such schemes are not common today [Liefke00].
[2] Actually, ASP and ASP.NET are implemented as ISAPI DLLs. However, the programming environment for ASP/ASP.NET is very different from ISAPI, thus we list it as a separate environment within the IIS system.

For TerraService, we define web service methods that take simple data-types as parameters, e.g. integers, strings, real numbers, and dates. The TerraService methods return the same list of simple data-types, an array of simple data-types, a struct of simple data-types or arrays of simple data-types, or a complicated struct of arrays, structs, or arrays of structs.

### 4.2 VisualStudio.NET Creates Web Services

Creating a web service in Visual Studio .Net is very easy. Run the development environment and create a *New Project*. One of the choices in the list of the Web Applications will be a *web service Project*, select it. The web service project will create a class definition for you that inherits from the System. System.Web.Services.WebService class. This gets you all the plumbing you need to marshal data to/from XML and send/receive SOAP messages.

Visual Studio .NET will create a project containing a set of source files. Two files distinguish a web service from any other type of .NET application program – the ".asmx" file and the ".asmx.cs" file.[3] The ".asmx.cs" file contains the Class definition statement that names our web service class and inherits from the System.Web.Services class. The ".asmx" file, as stated earlier, is a simple text file that is used by the ASP.NET run-time to locate the web service implementation class when invoked by a client SOAP request.

The web service class is implemented like any other public class. It has a parameter-less constructor that the ASP.NET environment calls on the first request to the web service. The class can perform any initial processing in this constructor. In the TerraService, the constructor formulates the connection strings required to access the TerraServer database and stores them in private variables to be used later by public methods.

Other class methods can be *private* and inaccessible to client applications or they can be *public* to consuming applications. Public methods that can be invoked using the SOAP protocol are decorated with the **[WebMethod]** attribute immediately preceding the declaration of the public method. The .NET Framework programming environment allows *annotations* to be recorded with compiled programs. These attributes can be used by other tools to enhance the program. In this case, the **[WebMethod]** attribute causes the framework to automatically generate the SOAP/XML documents that describe the published web service methods needed by consuming applications (the WSDL).

In summary, on Microsoft .NET platforms, a web service is a class containing public methods decorated with the **[WebMethod]** attribute. This attribute hides all the details of generating proxies and stubs, binding to the web service, and generating the metadata that clients need to invoke the web service. If they want to interoperate with all platforms web service authors are restricted to using a W3C standardized data-types and structures in the SOAP specification as method parameters and return values.

## 5 TerraService Web Service

The TerraService web service (http://TerraService.Net) provides a programmatic interface to the TerraServer database. A little background on the TerraServer database is required to understand the TerraService design and applications use it. Earlier research reports describe the Microsoft TerraServer database and HTML based application in depth [Barclay 98, Barclay 00]. Briefly, TerraServer is a database repository of high resolution, ortho-rectified imagery that is logically represented as a "seamless mosaic of earth" at several scales (meters per pixel). The TerraService also provides a landmark service that returns place names and a list of all the places within a specified area.

The mosaic is stored as sets of uniformly sized, JPEG or GIF compressed images called *tiles*. Each tile has a predictable resolution and location on the earth. The TerraService is a "tile service". TerraService methods enable consuming applications to query the TerraServer database of tiles by a number of different methods to determine the existence of tile data over some expanse of geographic territory. Data returned by tile query methods enable a calling application to determine the set of tiles required to build a complete image that covers the queried geographic area. The consuming application can iteratively call `TerraService's GetTile()` method to actually retrieve the individual tiles, construct a new, larger image, and then crop, re-size, and/or layer other graphical data on-top of the image.

In aerial imagery and remote-sensing, imagery is classified by the ground covered by an image pixel and by the number of color bands or samples captured for each pixel. The ground covered by a pixel is known as the image's resolution and is stated in meters per pixel "on a side", e.g. 1 meter per pixel, 2.54 meters per pixel, etc. The image band number and content of a pixel depends on the sensor that captured the image. Usually bands identify a color, red, green, blue, or grayscale. But a band can contain other data that might not even be visible normally, e.g. infrared. Each pixel in an USGS aerial image covers one square meter on the ground and comes in grayscale (black and white) or color-infrared (RGB).

TerraServer supports a powers-of-2 ladder of resolutions – .25, .50, 1, 2, 4, 8, 16, 32, through 16,384 meters per pixel. The 1 meter per pixel resolution is used as the "base resolution" in the system. All other resolutions are a multiple of 2 from 1 meter resolution.

All TerraServer tiles are 200 x 200 pixel images. A 1 meter resolution tile covers a 200 meter by 200 meter area. A 2 meter tile covers a 400 meter by 400 meter area, and so on.

---

[3] This article assumes the web service will be written in the C Sharp language. Web services can be implemented in any Visual Studio language. The Visual Basic equivalent would be the ".asmx.bas" file.

Typically, TerraServer stores a seven-level "image pyramid". The actual number of resolutions is dependent on the type of imagery being stored. TerraServer stores several mosaic *themes*. Currently two themes are accessible from the TerraService – aerial imagery (photograph), and topographic map (scanned from paper maps).

**USGS Digital Ortho-Quadrangles (DOQ)** images are gray-scale, 1-meter resolution aerial photos. Cars can be seen, but 1-meter resolution is too coarse to show people. Imagery is ortho-rectified to 1-meter square pixels. Approximately 90% of the U.S. has been digitized. The conterminous U.S. is expected to be completed by the end of 2002.

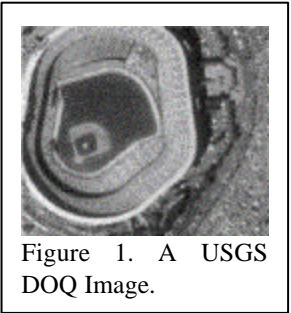

Figure 1. A USGS DOQ Image.

**USGS Digital Raster Graphics (DRG)** images are 13-color digitized topographic maps, with scales varying from 2.4 meter resolution to 25.6 meter resolution. DRGs are the digitized versions of the popular USGS topographic maps. The complete set of 65,000 USGS topographic maps have been scanned including Alaska, Hawaii, and several territories such as Guam and Puerto Rico.

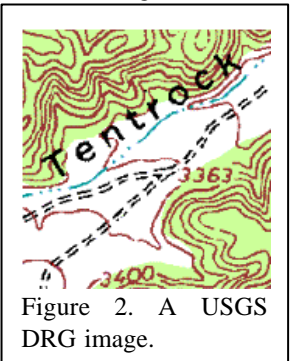

Figure 2. A USGS DRG image.

The most popular theme is the aerial imagery and it will be used as the primary example throughout this article.

Applications typically access the TerraService to build an image from multiple TerraServer tiles to use as a background image for some geo-spatial display. An application can use the meta-data returned by other TerraService methods to compute the ground coordinates in an image. This enables the applications to overlay additional text and graphics on the generated image. Figure 3 shows an image generated by a TerraService client application that added UTM grid lines.

### 5.1 About Map Projections

Each geographic system maintains spatial object coordinates. Over the centuries, many spatial coordinate systems have been developed. Each has unique properties that are designed to reduce the distortion of representing a sphere on a planar surface like a paper map or computer monitor screen.

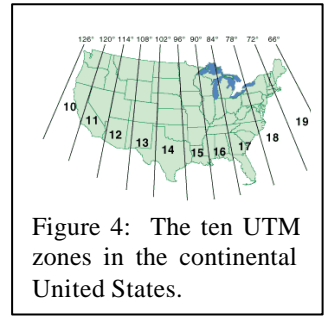

Figure 4: The ten UTM zones in the continental United States.

The USGS DOQ and DRG data stored in the TerraServer database use the Universal Transverse Mercator (UTM) projection. The ellipsoid of the projection is based on the North American Datum of 1983 (NAD83). Functions exist within the TerraService to convert from geographic coordinates (longitude and latitude) to and from UTM NAD83.

The UTM projection divides the earth into sixty 6° wide *UTM zones* sequentially numbered from 1 beginning at the International Date Line. Figure 4 shows the UTM zones that cover the conterminous United States. In the TerraServer, these zones are called *scenes*.

Applications that interact with the TerraService are required to understand the fundamental characteristics of the UTM projection system in order to be able effectively use TerraService data. For example, data from two different UTM zones cannot easily be combined together into a single map. Should an application want to build an image containing data from two zones, the application must be prepared to re-project the data of one or both of the UTM zones to build a new image.

In general, we expect TerraService applications to localize their use of TerraService to one UTM zone at a time. As you will see, the parameters to the TerraService methods and TerraService results assume that the calling application is working within one UTM zone at a time.

### 5.2 TerraService Methods

The TerraService web methods are divided into four categories – search, query tile meta-data, projection, and tile-fetch. The search, tile meta-data query, and projection methods are "helper" methods to help the application find the list of the tiles needed to build the desired image or map.

#### 5.2.1 Fetch Tile Method

Five parameters specify a tile's primary key in the TerraServer database.
- **Theme:** an enum selecting either an aerial image or a topo map tile.

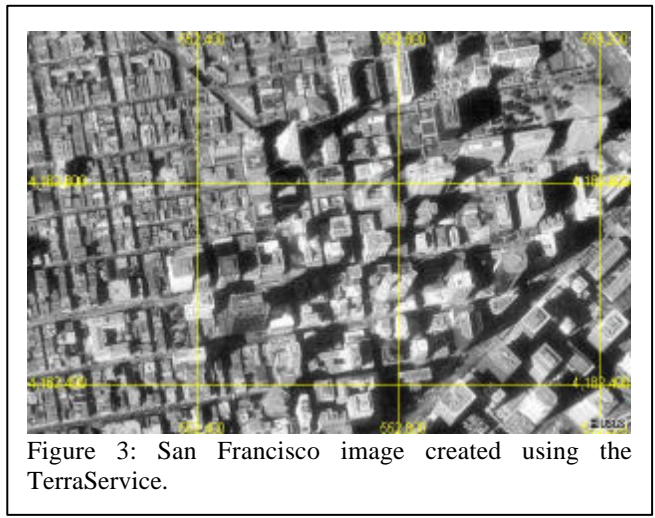

Figure 3: San Francisco image created using the TerraService.

- **Scale:** an enum that selecting the resolution the pixels resolution, e.g. 1 meter per pixel.
- **Scene:** an integer that identifies the seamless mosaic that the tile belongs too. For USGS aerial image and topo map data, the scene is the UTM Zone ( figure 4.)
- **X:** the tile's relative offset within the scene on the X-Axis. X values begin at 0 starting from the left or west side of the scene.
- **Y:** the tile's relative offset within the scene on Y-Axis. Y values begin at 0 starting from the bottom or south edge of the scene.

These five parameters make up the **TileId** struct parameter to the **GetTile()** web method. **GetTile()** returns an array of bytes that is the compressed tile in Jpeg or GIF format.[4] Adjacent tiles to a given tile can be determined by doing simple arithmetic to the **X** and **Y** fields of the TileId structure. The following table identifies the arithmetic to locate tiles immediate adjacent to the shaded tile below:

| X-1, Y+1 | X, Y+1 | X+1, Y+1 |
|---|---|---|
| X-1, Y | X, Y | X+1, Y |
| X-1, Y-1 | X, Y-1 | X+1, Y-1 |

### 5.2.2  *Tile Meta Data Methods*

The TerraServer database stores the longitude and latitude pair for each tile's corner points and center. The database also stores the date the image was photographed or published. The **GetTileMetaFromTileId()** method returns this information in a **TileMeta** struct that contains the **TileId** struct and also the tile's date and fiducial points.

Applications often want to locate a tile by geographic address. The **GetTileMetaFromLonLatPt()** method accepts a **LongLatPt** structure as a parameter and returns a **TileMeta** struct.

The **GetTileAreaFromPt()** web method describes the tiles required to build a map of a fixed image size with a specific Longitude and Latitude center point. The web method parameters are the desired image resolution, height, width, and center point. The method returns an **AreaBoundingBox** struct that contains five **AreaCoordinate** structs – one for each corner and one for the image center. An **AreaCoodinate** struct contains a **TileMeta** struct describing the tile that covers that corner or center of the image, and a **LonLatPtOffset** struct.

The **LonLatPtOffset** struct identifies the location of the corner or center pixel within a corner or center tile. For corner tiles it identifies where the image to be built should be cropped. The **LonLatPtOffset** of the center tile identifies the pixel offset of the absolute center of the image to be built. It also contains the **LonLatPt** struct that identifies the longitude and latitude of the corner or center point.

---
[4] USGS aerial imagery is returned in Jpeg format. 2 meter, 8 meter, and 32 meter topo map tiles are returned in GIF format. All other topo resolutions are returned in Jpeg format.

The **AreaBoundingBox** struct also contains the **NearestPlace** struct that identifies the name, direction to, and distance to the "nearest" significant landmark.

### 5.3  *Creating a Map using TerraService*

The Tile Meta Data methods and Fetch Tile methods are sufficient to build the map in Figure 3. It is a 600 pixel by 400 pixel rendering of downtown San Francisco using the *GetAreaFromPt()* method. The method was invoked using the following parameters:
- The Longitude and Latitude of the center of the image to be created
- The data Theme to use (USGS DOQ)
- The Scale of the image pixels (1 meter)
- The Width of the image (600)
- The Height of the image (400)

The *GetAreaFromPt()* method returned an **AreaBoundingBox** structure that contained the tile meta-data for each corner of the image and the image's center point. The loop in the code below calls the *GetTile()* method repeatedly to fetch and compose all the tiles required to build the image shown in Figure 3:

```
Int32 xStart = abb.NorthWest.TileMeta.Id.X;
Int32 yStart = abb.NorthWest.TileMeta.Id.Y;
Int32 xOff =(Int32)abb.NorthWest.Offset.XOffset;
Int32 yOff =(Int32)abb.NorthWest.Offset.YOffset;
for (Int32 x = xStart; x <=
            abb.NorthEast.TileMeta.Id.X; x++)
{
  for (Int32 y = yStart; y >=
            abb.SouthWest.TileMeta.Id.Y; y--)
  {
    TileId tid = abb.NorthWest.TileMeta.Id;
    tid.X = x;
    tid.Y = y;
    Image tileImage = Image.FromStream(
        new MemoryStream(ts.GetTile(tid)));
    compositeGraphics.DrawImage(tileImage,
      (x - xStart) * tileImage.Width - xOffset,
      (yStart - y) * tileImage.Height -yOffset,
      tileImage.Width, tileImage.Height);
    tileImage.Dispose();
  }
}
```

In the example, the **LonLatPtOffset** struct is use to compute how much to crop from the tiles on the image border. The **XOffset** value in the Northwest tile's **Offset** structure identifies the distance in pixels to crop from the left edge of the image. The **YOffset** value identifies the distance in pixels to crop from the top edge of the image.

The picture in Figure 3 shows yellow lines indicating the edges of the tiles that participated in forming the image. The lack of a yellow line on the picture edge indicates that the pixels were cropped from those edge tiles.

### 5.4  *Web Map Server*

The TerraService web methods are designed to support a OpenGIS compatible Web Map Server application. The OpenGIS is a consortium of organizations from the public and private sector interested in open standards for geo-spatial applications. Microsoft, the USGS, and USDA are members of the OpenGIS. The OpenGIS has published a standard for interoperable map applications on the Internet [OpenGIS].

The TerraService web site supports two map server applications – an OpenGIS Web Map Server (version 1.1.1) and a TerraService Map Server. The two map servers provide very similar functionality and differ primarily in the query string parameters used to generate a map. The OpenGIS map server can re-scale the image pixels whereas the TerraService Map Server does not offer image re-scaling services.

Both the OpenGIS Web Map Server and the TerraService Map Service were implemented using the TerraService web service class. Applications that access the TerraService web service class often use either map server to generate a single image from a set of TerraService tiles.

The OpenGIS conformant Web Map Server is accessible on the http://terraservice.net web site using the OGCWMS.ASHX web page. The following are the parameters for the OgcMap.ashx web page:

Table 1: OgcMap Parameters

| Param | Values | Description |
|---|---|---|
| Version | 1.1.1 | Required by OGC |
| Request | GetMap \| GetCapabilities | Info to return, a map or XML description of services supported |
| Layers | DOQ \| DRG | Type of map image to return on |
| Styles | blank \| UtmGrid \| GeoGrid | Grid lines to display on the image, default is none (blank) |
| SRS | EPSG:26903 – EPSG:26920 | Spatial Reference System of data. in UTM NAD83 zones 3 thru 20. |
| BBOX | minx#,miny#, max#,maxy# | Bounding Box coordinates of the image in UTM NAD83 values. |
| Width | 50 <= Pixel# <= 2000 | Width in pixels of resulting image |
| Height | 50 <= Pixel# <= 2000 | Height in pixels of resulting image |
| Format | image/jpeg | Mime type of the data to return |
| Exceptions | se_xml \| se_blank \| se_inimage | Error reporting style: – xml doc, blank image, or message in image |
| Service | wms | Required by Capabilities request |

Notes:

- **GetCapabilities** only requires Version, Request, and Service parameters. Other parameters only apply to Request=GetMap.

- The **Width** and **Height** parameters in combination with the Bounding Box parameters (BBOX) control the size of the image pixels produced. It is possible to generate images with pixel resolutions not stored within the TerraServer database, e.g. 11.375 meters per pixel.

The TerraService Map Server is accessible from http://terraservice.net/GetImageArea.ashx web page. The following are the parameters for the GetImageArea.ashx web page:

Table 2: TerraService Map Parameters

| Param | Values | Description |
|---|---|---|
| T | 1 (DOQ) \| 2 (DRG) | Data Theme to use as the base image |
| S | 10 – 16 | Scale of the pixels. 10 is 1 meters per pixel, 11 is 2 mpp, …and 16 is 64 mpp |
| Lon | Longitude degrees, | Center point of the image to produce |
| Lat | Latitude degrees, | Center point of the image to produce |
| W | 50 <= Pixels <= 2000 | Width in Pixels |
| H | 50 <= Pixels <= 2000 | Height in Pixels |
| F | Font name | Name of font for image text |
| FC | ARGB color string, | Hexadecimal color string, first two characters is the alpha blending value, next six characters is the RGB value |
| G | Pixel width | Width in pixels of grid lines, 0 = no grid |
| GC | ARGB color string | Grid Hexadecimal color string |
| B | Border width | Pixel width of image border |
| BC | ARGB color string | Boarder Hexadecimal color string |
| LOGO | 0 \| 1 | USGS logo in image if 1 |

## 6 Applications using TerraService

The United States Department of Agriculture (USDA), Natural Resources Conservation Service (NRCS) is responsible for programs to encourage sustainable management of natural resources. USDA is a cosponsor with USGS for the digital ortho-photography program.

The NRCS Information Technology Center (ITC) (www.itc.nrcs.usda.gov) developed two web-based applications that depend on the ortho-images provided by the TerraServer: the *Web Soil Data Viewer* and the *Environmental Easements Web Toolkit*. The Web Soil Data Viewer is a thin-client web application that displays color maps of interpreted soils data viewed against a background of ortho imagery obtained from the TerraServer. The composite map images are generated by Active Server Pages (ASP+) on the server and sent to the web browser. The Environmental Easements Web Toolkit is a rich-client web application that uses web services to directly obtain or "consume" ortho imagery from the TerraService and easement vector polygons from the ITC server which are separate server locations across the Internet.

### 6.1 Web Soil Data Viewer: a Server Side TerraService consumer

The Web Soil Data Viewer relies on four databases to interactively create interpretive soil maps and soil reports. The application integrates the soil geo-spatial data from two separate SQL Server databases. The TerraServer provides the digital ortho-photography and the ITC Lighthouse server creates the soils map using ArcIMS and the digital ortho image as a backdrop.

The Web Soil Data Viewer obtains soils information by using maps to navigate to an area of interest. The viewer sends an XML document describing the coordinates and desired data via the web server to the application server where ESRI ArcIMS is running.

ArcIMS parses the request and ArcSDE sends an index search on the data tables on SQL Server. ArcIMS returns the mapping tables via the IIS web server to the data viewer.

The next step is to zoom in to an appropriate image scale for the aerial photography. A custom COM+ middle-tier DLL retrieves tiles from the TerraService and builds the base image.

The format of the generated http request string is:
http://terraservice.net/GetImageArea.ashx?t=*T*&s=*S*&lon=*LON*&lat=*LAT*&w=*IW*&h=*IH*&f=Arial&fs=6&fc=ff000000

where **T** is set to 1 for DOQ imagery, **S** is set to 10 thru 16 for resolution, **Lon** is set to the center point's longitude, **Lat** is set to the center point's latitude, **IW** is set to the image width in pixels, and **IH** is set to image height in pixels.

TerraService replies with the ortho-image JPEG. Next, vector data such as roads, bridges and stream courses, are pulled from the SQL Server database and overlaid on the base ortho imagery to produce the final image sent back to the navigator.

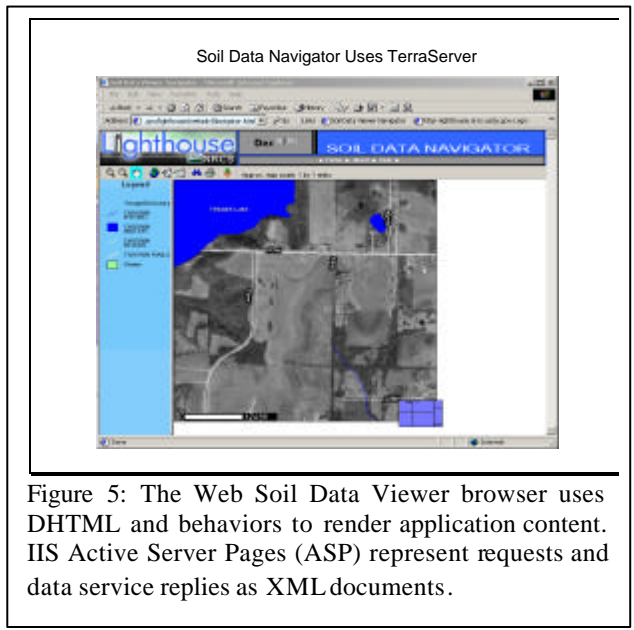

Figure 5: The Web Soil Data Viewer browser uses DHTML and behaviors to render application content. IIS Active Server Pages (ASP) represent requests and data service replies as XML documents.

### *6.2 Environmental Easements Application: a Client-Side Web Service Consumer*

NRCS works with local Conservation Districts to provide technical and financial incentives for landowners to implement conservation programs like the Wetlands Reserve Program (WRP) This program offers technical assistance and incentive to place private land into a wetlands restoration easement and restore or improve wetlands habitats. Biologists, Conservation Technicians, and District Conservationists use field visits, soils maps, and ortho-digital photograph for resource assessment and to develop wetlands restoration programs.

The Environmental Easements Toolkit was developed to support the Wetlands Restoration Program's easements tracking, location, status reviews, and as a geo-spatial collaborative work flow system. It shows the location of easements as alpha-blended tan polygons directly on a digital ortho-photo image obtained from the TerraServer web service (see Figure 6).

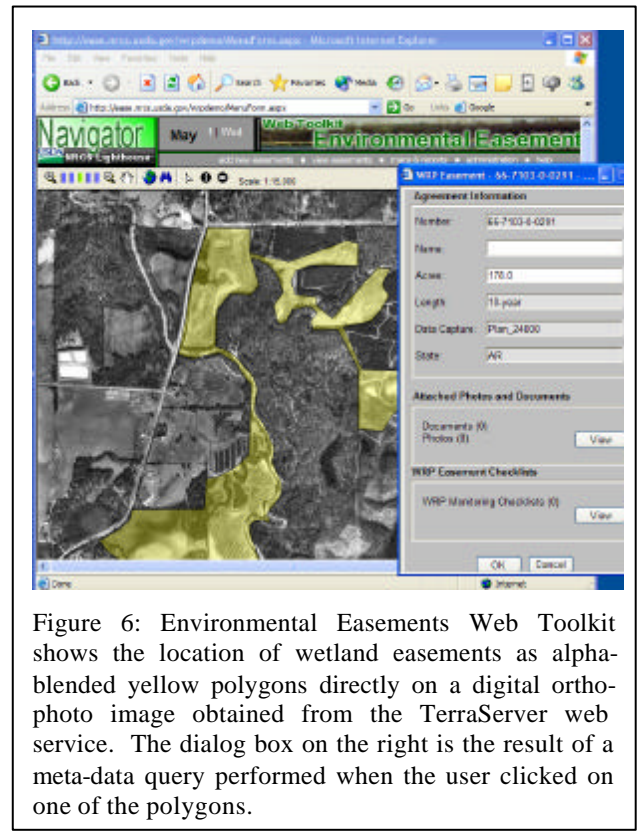

Figure 6: Environmental Easements Web Toolkit shows the location of wetland easements as alpha-blended yellow polygons directly on a digital ortho-photo image obtained from the TerraServer web service. The dialog box on the right is the result of a meta-data query performed when the user clicked on one of the polygons.

It is an intranet site which supports web-based digitization of TerraService images. Architecturally, the Environmental Easements application has three major components – Easements Client component, Easements web service, and TerraService. The client component accesses both web service applications and integrates data from them.

The Environmental Easements web service encapsulates geo-spatial tools of the ESRI Arc Spatial Data Engine (ArcSDE) [ArcSDE] that access USDA SQL databases. These methods are:
- **GetCentroids**: returns the polygon centroid.
- **LatLonExtentForOrtho**: returns the spatial extent for the orthophoto image in terms of latitude and longitude.

- **FindEasement**: given the easement number, returns the location data for the easement and for the subsequent invocation of TerraService.
- **GetEasement**: returns polygon data for the easement
- **GetEasementFromTerraServerArea**: returns polygon data for the easements within the extent of the TerraServer ortho image area
- **GetEasememntForUtmArea**: returns polygon data for the easements within the extent defined UTM coordinates
- **Update**: updates the easement databases
- **InsertGml**: converts and inserts the GML representation of the easement polygons

These web methods are used to re-project geospatial data, create and update attribute information associated with a spatial record, create a new spatial record, and generate XML for the geographic data. Geography Markup Language 2.0 (GML) is used to describe the polygons and the XML was extended to contain the parameters that access TerraService.NET [GML].

The Web Easement web service is stateless. Each method call passes all the context information required by the method to form a complete result. Like the TerraService, all parameters and method results use simple data-types such as strings, integers, and floating point values.

The client application first calls the *FindEasement()* method to locate an easement. The method takes three parameters – the SDELayer, the Easement Agreement number, and the display size to render. It returns an XML payload containing Geographic Markup Language 2.0 with the extended data elements for the TerraService parameters. It also returns the easement's polygons coordinates projected into the UTM zone coordinates.

The web browser client uses the values returned in the *FindEasement()* GML to fetch the TerraService jpeg tiles and display them in the browser. This is done with Jscript programs running in the Internet Explorer browser. The Jscript converts the GML payload from the FindEasement web service into Vector Markup Language and displays the easement polygons over the digital ortho imagery furnished by the TerraService. The client java script uses alpha-blending techniques so that the polygon is semi-transparent through. Figure 8 shows the final rendering of an easement displayed on top of an ortho-photographic image.

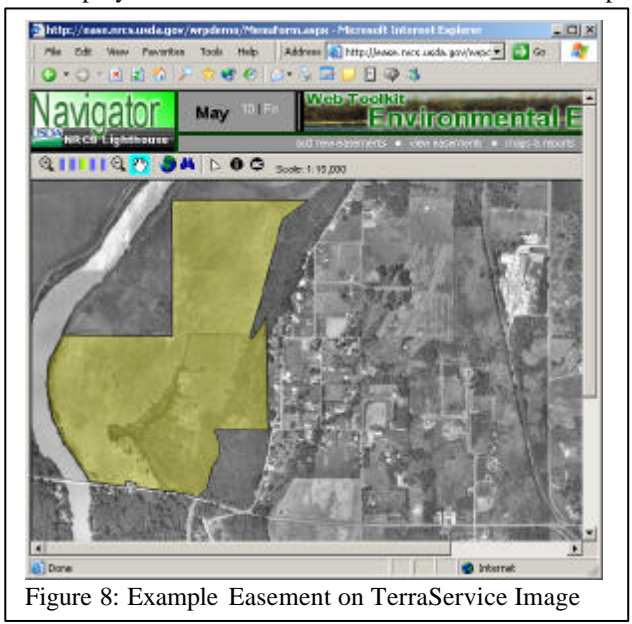

Figure 8: Example Easement on TerraService Image

### 6.3 Web Services as an Enabling Technology

The TerraService.NET web service provides geo-referenced digital ortho-photo images of the United States at several resolutions. It can be accessed programmatically through SOAP-based web services. The USGS digital ortho-images form an excellent base map for creating other data layers. Access to 4.5 terabytes of indexed image data provides an invaluable enabling technology for geospatial Internet application developers.

Before TerraService.Net the thought of using digital ortho-photo image data in web time with value added applications was not considered possible. With the imagery stored in a database, an image of any location in the United States is provided almost instantly.

The culmination of the distributed data processing between the federation of web services and client browser yields performance that creates a very acceptable user experience, even over low bandwidth network connections.

### 6.4 Solutions for Mission Critical Applications

Mission critical applications have stringent requirements for availability, stability, and performance. Web service based architectures can provide solutions that meet these requirements by enabling distribution and replication of data and processes to web service providers for asynchronous web-based computing.

Mission critical applications based on web services will require service agreements between the consumer and provider of web services to obtain success. Factors that should be considered are web service availability, performance, and redundancy. Duplication of web services placed in strategic network locations could provide failover, redundancy, or re-routing to ensure web service availability as in the case of network failure for a specific web service provider.

Web services offer developers a standards-based, cross-platform technology for building scalable, n-tier enterprise solutions and enabling Application Enterprise Integration. A web service approach to middle ware software development allows system designers the latitude to utilize heterogeneous operating systems and development environments. Web services, SOAP, and XML facilitate integration of disparate systems in the enterprise computing environment.

### 6.5 USDA Web-Service Applications

These two web-based applications show a range of application design choices: The Web Soil Data Viewer is a portal that integrates several web services at the web server and presents web pages to a thin client. The Environmental Easements Web Toolkit application is a rich-client design where the client performs the data integration of several web service data sources.

The deployed Environmental Easements Web Toolkit application owes much of its success to the pilot project implementation of the Web Soil Data Viewer. The Environmental Easements Web Toolkit will transition from limited operational pilot status to full operational production status in July 2002. NRCS users from across the country have quickly adopted this web-service based application to achieve their goal of creating and managing the national environmental easements geospatial database. NRCS is currently using this web-service architecture as a framework to design its next mission critical system that will provide much needed tools to accomplish the objectives of new Farm Bill.

## 7 Traffic & Performance

TerraService is supported by four web servers that connect to the TerraServer backend database server via ADO.NET. All the new web service code runs on these Compaq DL360s each containing two 700 mhz processors, 768 MB of RAM, and an 18 GB disk. The local disk stores the web service application software and the web logs. The four web servers were deployed in October 2000 in parallel with the Beta 1 release of Visual Studio.NET. Usage was low initially probably because only a few internet developers had access to Visual Studio.NET Beta 1.

Traffic to the TerraService web site increased steadily as more developers downloaded Visual Studio.NET Beta 2 in June 2001. Usage increased with the release of Visual Studio.NET in February 2002. Today several production applications use the TerraService facilities.

Since the TerraService was launched in October 2001, a total of 698,562 end users have visited the site or accessed an application that depends on the TerraService. The user volume is increasing. In June 2002, an average of 1400 users accessed the TerraService site per day during weekdays, and approximately 700 users per day on weekends. The peak users per day since February 2002 were 2023.

These users generate an average of 20,000 web requests during weekdays and about half as many requests on weekends. The peak web request volume since February 2002 was 130,886 which is over 6 times the normal volume. The web request volume on the TerraService site is divided into three major areas:

- 56% of web requests are the execution of the "Download USGS Image Application". This application uses the TerraService web service to construct multiple image tiles into a single Jpeg image with a USGS logo marked in the corner. The application is accessed from the Microsoft TerraServer end user web site (http://terraserver.microsoft.com).

- 14% of the web requests are direct calls to the TerraServer web service by remote applications.

- 5% of the web requests are calls to the TerraService Map Servers. Currently the USDA is the largest single user of the TerraService Map Server applications.

Even though this is a thin client implementation, the .NET web servers are lightly. We measured the average number of requests per second and processor load over a five hour period on a typical weekday. During the day, the site was visited by 1289 users that executed 21,087 web requests. Each web server handled an average of 6 web requests per minute with a maximum of 5 requests per second. The processor load ranged between 1.5% to a maximum of 9.5% on each web server. Thus, we have substantial overcapacity and are poised to handle a substantial increase in traffic.

## 8 Summary

The distributed systems community and vendors have long promised a workable scheme for building distributed applications. There was LU6.2, OLE, DSOM, DCOM, RMI, IIOP, CORBA, and many others. Each of these technologies worked in its own setting, but they did not interoperate, and they did work for Internet-scale applications.

It appears that XML Web Services is the first distributed object system that works at Internet-scale, is vendor neutral, and that has the tool suite that allows mere mortals to build, deploy, and operate applications. This is the convergence of solid web servers, XML for data formats, SOAP and WSDL for service definition and invocation, and a host of excellent integrated development environments from solid vendors.

The particular web services we have built use the Microsoft .NET™ toolkits and platforms. But the services present all their data in standard XML formats and so should interoperate with Macintosh and UNIX services.

In addition to the standard object-oriented rules for encapsulation and polymorphism there are some simple design rules that guide the design of a web service. If possible, make the methods atomic and stateless. Pick a granularity that lets each method run within a second and move less than 10KB over the network. Stick to XML that is widely supported by all the vendors and toolkits.

If one lives within these rules, it seems possible to quickly build scaleable and very functional web services that are easy to maintain, manage, and use.

## 9 References

[ArcSDE] ESRI Arc Spatial Data Engine http://www.esri.com/software/arcinfo/arcsde/index.html

[Barclay98] "The Microsoft TerraServer," Barclay, T., et. al., Microsoft Technical Report MS TR_98 17, Microsoft Corp, Redmond, WA.

[Barclay00] "Microsoft TerraServer: A Spatial Data Warehouse", Barclay, T., et. al., MS-TR-99-29. ACM SIGMOD2000, pp. 307-318

[Davis94] "EOSDIS Alternative Architecture Final Report," Davis, F., Sept., 1994, http://research.microsoft.com/~gray/EOS_DIS/

[Ekblad01] "Web Soil Data Viewer," Ekblad S., Strand E., and Aho, T., ESRI International User Conference, 2001.

[GeoML] Geography Markup Language 2.0 Specification, OGC Document Number: 01-029, http://www.opengis.net/gml/01-029/GML2.html

[Kobler95] "Architecture and Design of Storage and Data Management for the NASA Earth Observing System Data and Information System (EOSDIS)". Kobler B., et. al., : IEEE Symposium on Mass Storage Systems 1995: 65-76

[Liefke00] *"{XMill:} an efficient compressor for {XML} data,"* Liefke, H., Suciu, D., *Proc.* SIGMOD 2000, 153-164.

[Moore] Laurence Moore, "Transverse Mercator Projections and U.S. Geological Survey Digital Products", U.S. Geological Survey, Professional Paper.

[OMG] Object Management Group, http://www.omg.org/

[OpenGIS] Open GIS Consortium, http://www.opengis.org/

[Robinson95] *Elements of Cartography, Sixth Edition,* Robinson, A.H., et. al., John Wiley & Sons, Inc., U.S.A. 1995, ISBN 0-471-55579-7.

[Samet90] *The Design and Analysis of Spatial Data Structures,* H. Samet, Addison-Wesley, Reading, MA, 1990. ISBN 0-201-50255-0.

[Snyder89] *An Album of Map Projections*, Snyder, J.P., U.S. Geological Survey, Professional Paper, 1453, (1989).

[SQL Server] Microsoft SQL Server 7.0 http://microsoft.com/SQL/

[W3C] The World Wide Web Consortium, http://www.w3.org/ and especially the XML, SOAP, and web services directories.